\documentclass[10pt,conference]{IEEEtran}
\IEEEoverridecommandlockouts
\usepackage{dirtree,array}
\usepackage{tikz}
\usepackage{minted}
\usepackage{soul, color, xcolor}
\usepackage{amsmath,amsfonts}
\usepackage{algorithmicx}
\usepackage{graphicx}
\usepackage{textcomp}
\usepackage{xcolor}
\usepackage{fontawesome}
\usepackage{siunitx}
\usepackage{comment}
\usepackage{tcolorbox}
\usepackage{listings}
\usepackage{xspace}
\usepackage{cancel}
\usepackage{float}
\usepackage{graphicx}
\usepackage{subfigure}
\usepackage{pifont}
\usepackage{threeparttable,booktabs,multirow,lscape}
\usepackage{multirow}
\usepackage{adjustbox}
\usepackage{enumerate}
\usepackage{listings}
\usepackage{tabularx}
\usepackage[ruled,vlined]{algorithm2e}
\usepackage{amsmath}
\usepackage{environ}
\usepackage{tikz}
\usepackage{color}

 \usepackage{amssymb}
\usepackage{enumitem}
\usepackage{rotating}
\usepackage{threeparttable}
\usepackage{url}
\usepackage{hyperref}
\usepackage{bbding}
\usepackage{caption}
\usepackage{threeparttable}
\usepackage{listings, xcolor}

\definecolor{verylightgray}{rgb}{.97,.97,.97}

\lstdefinelanguage{Solidity}{
	keywords=[1]{anonymous, assembly, assert, balance, break, call, callcode, case, catch, class, constant, continue, constructor, contract, debugger, default, delegatecall, delete, do, else, emit, event, experimental, export, external, false, finally, for, function, gas, if, implements, import, in, indexed, instanceof, interface, internal, is, length, library, log0, log1, log2, log3, log4, memory, modifier, new, payable, pragma, protected, pure, push, require, return, returns, revert, selfdestruct, send, solidity, storage, struct, suicide, super, switch, then, throw, true, try, typeof, using, value, view, while, with, addmod, ecrecover, keccak256, mulmod, ripemd160, sha256, sha3}, % generic keywords including crypto operations
	keywordstyle=[1]\color{blue}\bfseries,
	keywords=[2]{MAX\_UINT, bool, byte, bytes, bytes1, bytes2, bytes3, bytes4, bytes5, bytes6, bytes7, bytes8, bytes9, bytes10, bytes11, bytes12, bytes13, bytes14, bytes15, bytes16, bytes17, bytes18, bytes19, bytes20, bytes21, bytes22, bytes23, bytes24, bytes25, bytes26, bytes27, bytes28, bytes29, bytes30, bytes31, bytes32, enum, int, int8, int16, int24, int32, int40, int48, int56, int64, int72, int80, int88, int96, int104, int112, int120, int128, int136, int144, int152, int160, int168, int176, int184, int192, int200, int208, int216, int224, int232, int240, int248, int256, mapping, string, uint, uint8, uint16, uint24, uint32, uint40, uint48, uint56, uint64, uint72, uint80, uint88, uint96, uint104, uint112, uint120, uint128, uint136, uint144, uint152, uint160, uint168, uint176, uint184, uint192, uint200, uint208, uint216, uint224, uint232, uint240, uint248, uint256, var, void, ether, finney, szabo, wei, days, hours, minutes, seconds, weeks, years},	% types; money and time units
	keywordstyle=[2]\color{teal}\bfseries,
	keywords=[3]{block, blockhash, coinbase, difficulty, gaslimit, number, timestamp, msg, data, gas, sender, sig, value, now, tx, gasprice, origin},	% environment variables
	keywordstyle=[3]\color{violet}\bfseries,
	identifierstyle=\color{black},
	sensitive=true,
	comment=[l]{//},
	morecomment=[s]{/*}{*/},
	commentstyle=\color{gray}\ttfamily,
	stringstyle=\color{red}\ttfamily,
	morestring=[b]',
	morestring=[b]"
}

\def\BibTeX{{\rm B\kern-.05em{\sc i\kern-.025em b}\kern-.08em
    T\kern-.1667em\lower.7ex\hbox{E}\kern-.125emX}}
\begin{document}

\title{Esim: EVM Bytecode Similarity Detection Based on Stable-Semantic Graph}

\author{\IEEEauthorblockN{Zhuo Chen}
% \IEEEauthorblockA{\textit{Zhejiang Univercity} \\
% hypothesiser.hypo@zju.edu.cn}
\and
\IEEEauthorblockN{Gaoqiang Ji}
% \IEEEauthorblockA{\textit{Zhejiang Univercity} \\
% jigaoqiang@zju.edu.cn}
\and
\IEEEauthorblockN{Yiling He}
% \IEEEauthorblockA{\textit{University College London} \\
\and
\IEEEauthorblockN{Lei Wu}
% \IEEEauthorblockA{\textit{Zhejiang Univercity} \\
% lei\_wu@zju.edu.cn}
\and
\IEEEauthorblockN{Yajin Zhou}
% \IEEEauthorblockA{\textit{Zhejiang Univercity} \\
% yajin\_zhou@zju.edu.cn}
}

\maketitle

\newcommand{\tab}{\hspace*{1em}}
\newcommand{\greenbullet}{\tikz\draw[green,fill=blue] (0,0) circle (.5ex);}
\newcommand{\redbullet}{\tikz\draw[red,fill=red] (0,0) circle (.5ex);}

\newcommand{\code}[1]{{\fontfamily{cmtt}\fontseries{m}\fontshape{n}\selectfont\small{#1}}}
\newcommand*{\prompt}[1]{\textsf{\textbf{#1}}}
\newcommand{\etc}{\emph{etc}\xspace}
\newcommand{\ie}{\emph{i.e.}\xspace}
\newcommand{\eg}{\emph{e.g.}\xspace}
\newcommand{\Ie}{\emph{I.e.}\xspace}
\newcommand{\Eg}{\emph{E.g.}\xspace}
\newcommand{\system}{\textit{Esim}}
\newcommand{\target}{UEware}
\newcommand{\legit}{normal app}
\newcommand{\leg}{normal}
\newcommand{\hypo}[1]{\textcolor{purple}{[hypo: #1]}}
\newcommand{\qy}[1]{\textcolor{purple}{[hypo: #1]}}
\newcommand{\lei}[1]{\textcolor{red}{[lwu: #1]}}
\newcommand{\sfg}{SSG}
\newcommand{\kcfg}{SCFG}
\newcommand{\kdfg}{SDFG}
\newcommand{\solc}{Solc}
\newcommand{\sinst}{stable instruction}

\begin{abstract}

Decentralized finance (DeFi) is experiencing rapid expansion. However, prevalent code reuse and limited open-source contributions have introduced significant challenges to the blockchain ecosystem, including plagiarism and the propagation of vulnerable code. 
Consequently, an effective and accurate similarity detection method for EVM bytecode is urgently needed to identify similar contracts. %, thereby facilitating plagiarism detection and vulnerability assessment.
Traditional binary similarity detection methods are typically based on instruction stream or control flow graph (CFG), which have limitations on EVM bytecode due to specific features like \textit{low-level} EVM bytecode and \textit{heavily-reused} basic blocks. 
% Low-level bytecode disrupts stream-based instruction approaches that treat binary code as natural languages. The instability of basic blocks results in the failure of CFG-based approaches that rely on the relation between basic blocks. 
% In addition, the \textit{highly-diverse} Solidity Compiler (Solc) version also increases the challenge of similarity detection.
Moreover, the \textit{highly-diverse} Solidity Compiler (Solc) versions further complicate accurate similarity detection.

Motivated by these challenges, we propose a novel EVM bytecode representation called \textit{Stable-Semantic Graph} (SSG), which captures relationships between ``stable instructions'' (special instructions identified by our study).
Moreover, we implement a prototype, \system{}, which embeds SSG into matrices for similarity detection using a heterogeneous graph neural network.
\system{} demonstrates high accuracy in SSG construction, achieving F1-scores of 100\% for control flow and 95.16\% for data flow, and its similarity detection performance reaches 96.3\% AUC, surpassing traditional approaches. 
% Subsequently, we analyzed 2,675,573 contracts across 6 EVM chains over one year. 
% The average SSG construction time is 26s, and the average similarity search time is less than 5s.
Our large-scale study, analyzing 2,675,573 smart contracts on six EVM-compatible chains over a one-year period, also demonstrates that \system{} outperforms the SOTA tool Etherscan in vulnerability search.

\end{abstract}

% \begin{IEEEkeywords}
% bytecode similarity, EVM
% \end{IEEEkeywords}

\section{Introduction}

With the rapid expansion of decentralized finance (DeFi) in the blockchain ecosystem, DeFi projects, which are built on smart contracts on the Ethereum Virtual Machine (EVM), have attracted substantial investment in recent years, with over \$88.8 billion Total Value Locked (TVL) in 2024~\cite{Defillma}. 

However, extensive code reuse and the rarity of open-source code are prevalent in the blockchain ecosystem~\footnote{More than 99\% of the Ethereum contracts are not open source~\cite{Etherscan}}. 
% many projects with large TVLs fork top DeFi protocols~\cite{DefillmaForks}. 
These phenomena raise serious concerns for DeFi projects, such as plagiarism and propagation of vulnerable code. 
% When vulnerabilities in popular protocols or components are exploited, the forked projects are also exposed to attacks. 
As a representative case, the Compound v2 protocol~\cite{Compound}, one of the top lending protocols, has been widely adopted and forked by numerous DeFi projects. 
This protocol has a known precision loss issue that can be exploited when the corresponding market lacks liquidity.
Since 2022, a series of attacks (\eg, Hundred Finance Attack~\cite{hundredfinance}, Onyx Protocol Attack~\cite{OnyxProtocol}, Radiant Attack~\cite{Radiant}) have been observed due to the code abuse of Compound v2 protocol, resulting in millions of dollars in losses. 
% The most recent such attack occurred in May 2024~\cite{Sonne}.
Consequently, there is an urgent need for an efficient method to detect code reuse in EVM bytecode (binaries), a process also known as EVM bytecode similarity detection.
% Specifically, binary code similarity detection is a fundamental task that determines the semantic similarity between two binary functions, which have proven to be effective in addressing the above concerns in traditional domains, \eg, C++~\cite{xu2017neural,he2024code,feng2016scalable} and Java~\cite{dann2019sootdiff}. 

In general, binary similarity detection studies in traditional languages (\eg, C++~\cite{xu2017neural,he2024code,feng2016scalable} and Java~\cite{dann2019sootdiff}) can be divided into two categories, \ie, instruction stream based and control flow graph (CFG) based. 
\textbf{Instruction stream based studies} treat instructions as natural language sentences and introduce natural language processing (NLP) techniques~\cite{zuo2018neural,pei2022learning,wang2022jtrans}. 
% The methods in this direction leverage the specially designed pre-training tasks and datasets to enable the deep neural network model to grasp code semantics.
\textbf{CFG based studies} treat basic blocks as the smallest logical unit, utilizing control flow information. In some cases, researchers have also incorporated data flow information into basic block nodes to enhance analysis capabilities~\cite{feng2016scalable,xu2017neural}.

Previous studies have encountered three main challenges with EVM binaries:
\noindent \textbf{(Challenge I)} \textit{Low-level EVM bytecode.} Compared to high-level bytecode (\eg, Java), the semantic loss of EVM bytecode is more pronounced, lacking symbolic information and having only low-level branches (\ie, \code{jump}).
The semantic disparity leads to the low effectiveness of instruction flow based methods, which heavily rely on well-structured language.
\noindent \textbf{(Challenge II)} \textit{Heavily-reused basic block.}
Considering the gas consumption and code size limit in EVM, the Solidity compiler (Solc) incorporates a basic block reuse strategy. This strategy aims to reduce the size of the bytecode by breaking down the original semantically complete basic blocks into multiple smaller ones. In this way, other functions can reuse these smaller basic blocks.
As a result, the control flow semantics among basic blocks is disrupted, resulting in an unclear function boundary and incorrect basic block relationships. It leads to the failure of previous CFG-based studies on basic block granularity.
In addition to domain-specific challenges, there is a common challenge that must be addressed. 
\noindent \textbf{(Challenge III)}  \textit{Highly-diverse Solc versions.} \solc{} has developed rapidly, with a total of 113 versions~(from 0.1.3 to 0.8.24) in 8 years. The Solc in different compilation versions and options is very different, resulting in huge differences in the bytecode even for the same source code.

% These challenges indicate that traditional methods in other languages are ineffective in EVM bytecode. 
Few studies are dedicated to EVM bytecode similarity detection~\cite{liu2018eclone, he2020characterizing}. However, their EVM bytecode representations still refer to the traditional ones used in other languages, lacking improvements tailored to EVM bytecode. 
% To address these challenges, we suggest the development of an EVM bytecode representation that is capable of \textit{(i)} effectively connecting low-level bytecode instructions with high-level semantics while excluding elements irrelevant to semantics, such as error handling. 
% \textit{(ii)} handling the unclear function boundaries and heavily basic blocks reuse strategy of \solc{}.
% \textit{(iii)} maintaining stability across different optimization and compilation versions, thereby ensuring excellent coverage (see details in Section~\ref{subsec:insight}). 
In this paper, we propose a novel EVM bytecode representation, named \textbf{Stable-Semantic Graph (SSG)}. 
% The \sfg{} consists of the control flow and data flow information among the specific instructions, \ie, storage, log, call, and return instructions, which we refer to as \textbf{high-semantic instructions} in this paper. This new representation is based on two key insights (Section~\ref{sec:background}) derived from the \textit{stable} instructions: 
The \sfg{} consists of the control flow and data flow information among the specific instructions, \ie, storage, log, call, and return instructions, which is designed based on two key insights (Section~\ref{sec:background}) derived from the \textbf{stable instructions}: 
\begin{itemize}[leftmargin=*]
\item \noindent\textit{EVM instructions have graded semantic relevance.} 
The EVM dictates that solely storage-related instructions are capable of altering the global state of the blockchain. Conversely, stack-related instructions are only able to modify the information within the template stack.
When implementing a DeFi project, functions interact and cooperate via call-related instructions. Additionally, log-related instructions are employed to facilitate off-chain analysis. These instructions exert a more substantial influence on the blockchain, thereby having high semantic relevance.

\item \noindent\textit{Instructions with high semantic relevance are more stable against the basic block reuse strategies.}
Instructions with high semantic relevance are of greater significance and cost more than those with low semantic relevance (i.e., those related to stack or memory operations). To ensure program correctness, Solc will reduce the reuse of basic blocks where these instructions are located, thereby maintaining stable connectivity and order.

% According to the high DeFi semantic relevance between these instructions, these instructions are more expensive and important than those related to stack or memory operations. This makes them less prone to basic block reuse, maintaining a stable connection and order.
\end{itemize}

Drawing on our EVM bytecode representation, we implement a prototype, \system{}, for detecting the similarity of EVM bytecode.
Specifically, \system{} consists of two modules: \textit{(i) \sfg{} builder}, which constructs the \sfg{}s of a given EVM contract bytecode at the function level. 
% This module consists of two sub-processes, \ie, stable control flow graph construction and stable data flow graph integration. The \sfg{} builder finally outputs the \sfg{}s.
\textit{(ii) \sfg{} embedding generator}, which is responsible for extracting features from the \sfg{} and transforming it into a vector for similarity detection. 
% Specifically, to better represent the control flow and data flow information, we leverage a heterogeneous graph embedding model.

To evaluate \system{}, we build up three datasets. 
\begin{itemize}[leftmargin=*]
    \item \textit{Dataset I}: Source code and corresponding bytecode from two famous open-source DeFi protocols (\ie, Uniswap V2, USDT). This dataset is used to evaluate the \sfg{} construction accuracy.
    \item \textit{Dataset II}: Contracts bytecodes compiled from the same source code using multiple versions of Solc~\footnote{Given that these bytecodes were compiled from the identical source code, we possess the ground truth of these bytecodes. Consequently, we utilize this dataset for both training and evaluation.}. 
    % Specifically, when compiling the same source code with different Solc versions, we manually inspected the syntax changes induced by various compilation versions and corrected these syntax errors based on the \solc{} Abstract Syntax Tree (AST). 
    We selected the top 10,000 Total Value Locked (TVL) Ethereum projects (by DeFillma~\cite{Defillma}) and obtained 21,018 \sfg{}s representing different functions. This dataset is used for both training and detection performance evaluation.
    
    % Specifically, to compile the same source code in different Solc versions, we manually reviewed the syntax changes caused by various compilation versions and adjusted these syntax errors based on the \solc{} Abstract Syntax Tree (AST). Finally, we choose the top 10,000 Ethereum projects, and get 21,018 \sfg{}s of different functions.
    \item \textit{Dataset III}: Contracts bytecodes deployed on six EVM-compatible chains~\footnote{Including Ethereum~\cite{Ethereum}, Optimisim~\cite{Optimism}, Binance Smart Chain (BSC)\cite{Binance}, Polygon POS\cite{Polygon}, Arbitrum One\cite{Arbitrum}, and Avalanche-C-Chain\cite{Avax}.} between January 1st, 2023 and June 13th, 2024. In total, we collect 2,675,573 distinct contracts and their corresponding bytecode. This dataset is used to evaluate the efficiency and real-world performance of \system{}.
\end{itemize}

% First, we use two famous source-available DeFi protocols as \textit{Dataset I} to evaluate the \sfg{} construction accuracy.
The evaluation results indicate that the construction of \sfg{} is highly accurate. Specifically, \system{} achieves a 100\% F1-score in control flow identification and 95.16\% F1-score in data flow identification.
In the realm of similarity detection performance, \system{} outperforms other studies~\cite{he2020characterizing,liu2018eclone,feng2016scalable,Etherscan} and reaches the state-of-the-art (SOTA) level. It attains \textbf{an AUC~\footnote{A common metric for binary classification: Area Under the receiver operating characteristic Curve.} of 96.3\%}, which is at least \textbf{12.6\% higher} than that of previous research. Meanwhile, our evaluation verifies that our system is insensitive to hyperparameters, as they only influence the AUC effectiveness by 2.8\%.
In terms of efficiency, the average time required to extract \sfg{}s from a single contract is 26 seconds, and the time taken for similarity detection across six EVM-compatible chains is less than 5 seconds.

Finally, we conducted a real-world downstream task, \ie, vulnerability code search to evaluate \system{}.
Our evaluation results indicate that \system{} outperforms Etherscan~\cite{Etherscan}, the most commonly used similarity detection tool. For vulnerability codes attacked between May 2023 and May 2024, \system{} successfully identified 382 similar vulnerable contracts without any false positives.
In stark contrast, Etherscan only identified 255 similar contracts, and the false positive rate of its results even exceeded 93.3\%.

% The result shows that \system{} significantly outperforms the most widely adopted similarity detection tool, Etherscan. 
% For vulnerability codes attacked between May 2023 and May 2024~\footnote{We only focus on attacks resulting in losses exceeding \$1 million}, we identified 382 similar vulnerable contracts with no false positives. In contrast, Etherscan identified only 255 authentic similar vulnerable contracts, with an over 93.3\% false positive rate.

% These case studies demonstrate that \system{} is highly effective in real-world scenarios, helping to mitigate user losses by identifying malicious patterns.

\noindent \textbf{Contributions.}\tab
We make the following contributions:
\begin{itemize}[leftmargin=*]
\item We propose a novel representation of EVM bytecode, named stable-semantic graph (SSG), to effectively perform similarity detection.
\item We design and implement a prototype, \system{}, and perform extensive experiments and demonstrate the effectiveness of \system{} over previous studies.
\item We conduct large-scale real-world case studies involving 2,675,573 contracts, demonstrating the effectiveness of \system{} in vulnerability code search.
\item The prototype of the proposed \system{} system is currently available in an anonymized repository~\footnote{Our repository in \url{https://anonymous.4open.science/r/esim-CF8D}}. Both the prototype and its accompanying dataset will be made publicly available upon acceptance.
\end{itemize}

\section{Background \& Motivation}
\label{sec:background}

In this section, we first illustrate the background of the EVM bytecode (Section~\ref{subsec:toy}), and then discuss the challenges and distinctions encountered when employing traditional representation techniques for the EVM bytecode (Section~\ref{subsec:challenge}).
Finally, we present our insights (Section~\ref{subsec:insight}) and show the formal definition of \sfg{} (Section~\ref{subsec:defination}).

\subsection{Characteristics of the EVM Bytecode}
\label{subsec:toy}
\begin{figure}[!t]
\centering
\includegraphics[width=.45\textwidth]{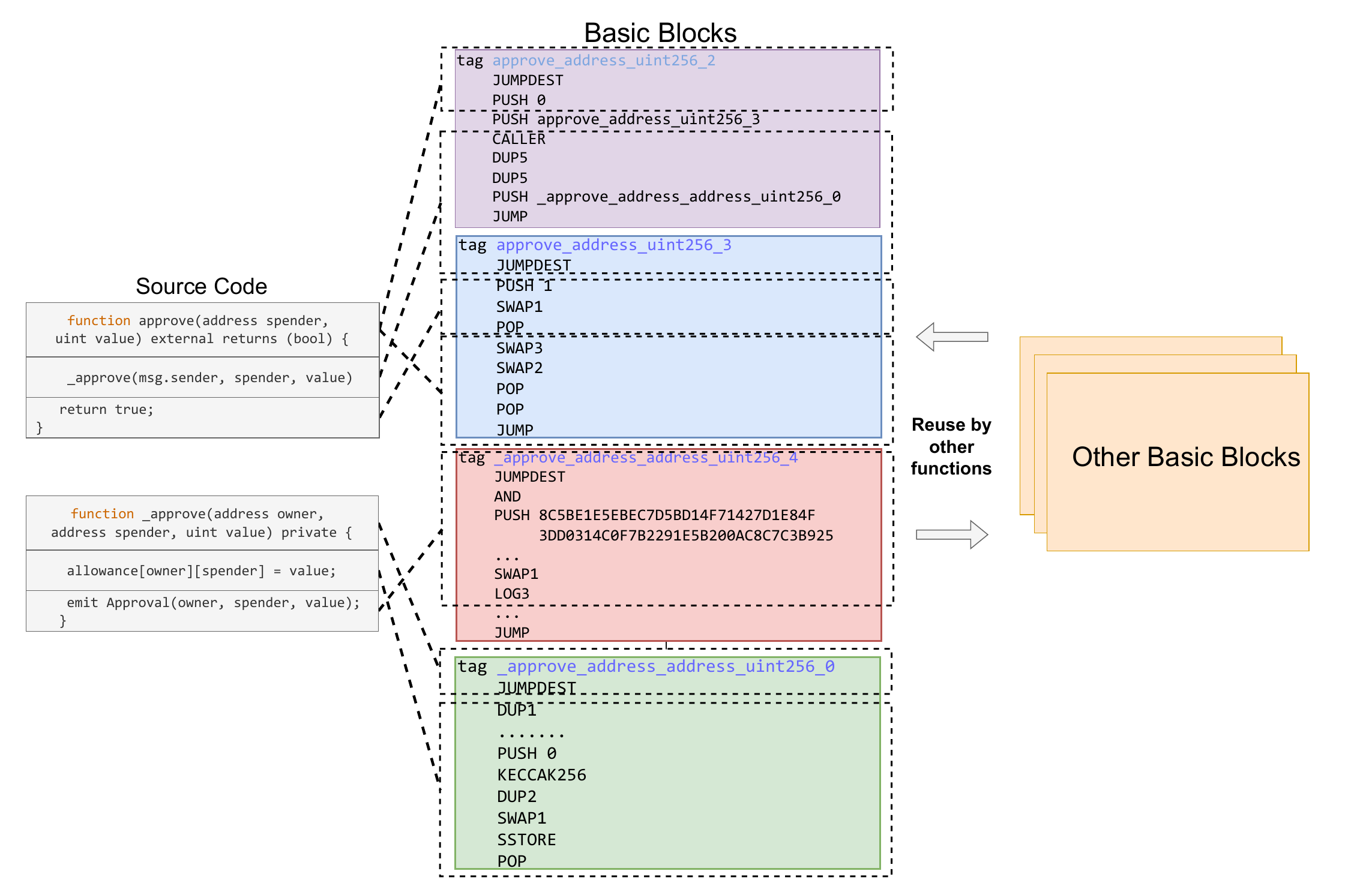}
	\caption{A simple EVM bytecode example. Due to the basic block reuse strategy employed by \solc{}, the non-branching source code will be divided into multiple basic blocks.}
	\label{fig:simplebinary}
        \vspace{-1em}
\end{figure}

% We show a simple \code{approve} function. As shown in Figure~\ref{fig:simplebinary}, 
Figure~\ref{fig:simplebinary} shows a simple \code{approve} function in Solidity.
This external \code{approve} function calls the private \code{\_approve} function and returns true, where the \code{\_approve} function changes the storage \code{allowance} according to the \code{calldata} from the caller and \code{emit} a log, with the \code{owner, spender}, and \code{value} variables.

% In traditional languages, using Java as an example, both the private and the external functions are registered in the function table as two individual functions. Due to the absence of any branching statements, these two functions would be compiled into two basic blocks. 
In many traditional compiled languages, like Java, the external \code{approve} function and the private \code{\_approve} function would typically be compiled as two separate functions. If these functions contain no internal branching statements, each might also be compiled into a single basic block. 

% However, the EVM bytecode is different. 
The Solidity compiler (\solc{}) handles this situation quite differently when generating EVM bytecode.
Firstly, \solc{} inlines the private function into the external function, resulting in an unclear function boundary.
Secondly, the external and private functions are compiled into four basic blocks instead of two. This is due to the block reuse strategy of \solc{}, which allows a part of the code of these functions to be reused by other functions. Consequently, \solc{} separates the complete logic of \code{approve} using jumps into multiple smaller basic blocks, so that the smaller basic blocks can be reused by other functions and thus reduce the contract size.

\subsection{Problem Challenges}
\label{subsec:challenge} 
% Over the past few years, many studies have encountered code similarity problems in other languages, such as Java and C++. 
The techniques for solving code similarity problems can be classified into two directions, \ie, instruction stream based and control flow graph (CFG) based.
\textbf{Instruction stream based studies} treat instructions as natural language sentences and introduce natural language processing (NLP) techniques~\cite{zuo2018neural,pei2022learning,wang2022jtrans}.
\textbf{CFG based studies} treat basic blocks as the smallest logical unit, utilize control flow information to judge the similarity. In some cases, researchers incorporate data flow information into basic block nodes to enhance analysis capabilities~\cite{feng2016scalable,xu2017neural}.

However, traditional studies are incompatible with EVM bytecode and face the following challenges:
\begin{itemize}[leftmargin=*]
    \item \textbf{Low-level EVM Bytecode.} 
    Compared to high-level bytecode (\eg, JVM), semantic information loss is far more severe in EVM bytecode, complicating the extraction of semantics. Notably, symbolic information is absent in EVM bytecode. For example, EVM bytecode lacks construct and name definitions for objects and methods. Furthermore, all method calls rely solely on the \code{jump} and \code{jumpi} instructions, further obfuscating the code's semantic meaning.
    
    As depicted in Figure~\ref{fig:simplebinary}, both function/variable names and type definitions are missed after compilation (the function tag names are our annotations for better understanding).
    It disrupts instruction flow-based research that heavily depends on well-structured, high-semantic codes.
    
    \item \textbf{Heavily-reused Basic Block.}
    \solc{} adopts a basic block reuse strategy to minimize bytecode size. 
    Specifically, it splits the original semantically complete basic blocks into several smaller ones. Consequently, other functions can reuse these smaller basic blocks, thereby achieving the objective of optimizing bytecode size.
    
    The basic blocks reuse strategy leads to the following problems:
    \textit{(i) Function boundaries are unclear.} 
    The fragmentation of basic blocks into smaller, reusable ones blurs the lines between functions. As a result, discerning function boundaries by analyzing the connectivity among these small basic blocks is difficult. In Figure~\ref{fig:simplebinary}, the \code{\_approve} function is split into two basic blocks, and it is hard to determine its boundary from bytecode.
    \textit{(ii) Basic Block control-flow is obfuscated.} The basic block reuse strategy affects the order of basic blocks. Thus, the program's execution order depends on the stack value read before the \code{jump} instruction, blurring the line between data-flow and control-flow analysis in EVM static analysis.
    In Figure~\ref{fig:simplebinary}, the order of the basic blocks in the \code{\_approve} function is swapped due to the reuse strategy.
    Consequently, traditional CFG based studies are not effective in EVM bytecode.

    \item \textbf{Highly-diverse Solc Versions.} 
    The \solc{} has developed rapidly, with a total of 113 versions from 0.1.3 to 0.8.24 in 8 years. 
    % The frequent version updates have caused a phenomenon: the DeFi project team can freely choose different compiler versions/optimization options to compile source code into bytecode.
    However, the \solc{} in different compilation versions are very different, resulting in huge differences in the bytecode even for the same source code. 
    Simultaneously, within the same version, the optimization level of \solc{} also affects the EVM bytecode. At higher optimization levels, \solc{} tends to more frequently divide basic blocks.
    
    % The major differences are as follows: \textit{(i) Condition checking and error handling.} As the version changes, the condition checking and error handling routine has a large gap, which leads to confusion about the same code. 
    % \textit{(ii) Optimization level.} The \solc{} optimization level greatly affects EVM bytecode. For higher optimization, \solc{} will frequently split a logical basic block and reuse its paragraphs. This can lead to huge differences in the control flow and basic blocks between low and high optimization levels.
    
\end{itemize}

% In addition to domain-specific challenges, there exists a common challenge that must be addressed to ensure comprehensive detection coverage.
% \begin{itemize}
    % \item \textbf{Highly-diverse Solc Versions.} 
    % The \solc{} has developed rapidly, with a total of 113 versions from 0.1.3 to 0.8.24 in 8 years. 
    % The frequent version updates have caused a phenomenon: the DeFi project team can freely choose different compiler versions/optimization options to compile source code into bytecode. However, the \solc{} in different compilation versions and options are very different, resulting in huge differences in the bytecode even for the same source code. 
    % The major differences are as follows: \textit{(i) Condition checking and error handling.} As the version changes, the condition checking and error handling routine have a large gap, which leads to the confusion of the same code. 
    % \textit{(ii) Optimization level.} The \solc{} optimization level greatly affects EVM bytecode. For higher optimization, \solc{} will frequently split a logical basic block and reuse its paragraphs. This can lead to huge differences in the control flow and basic blocks between low and high optimization levels.
% \end{itemize}

\begin{figure}[!t]
\centering
\includegraphics[width=.45\textwidth]{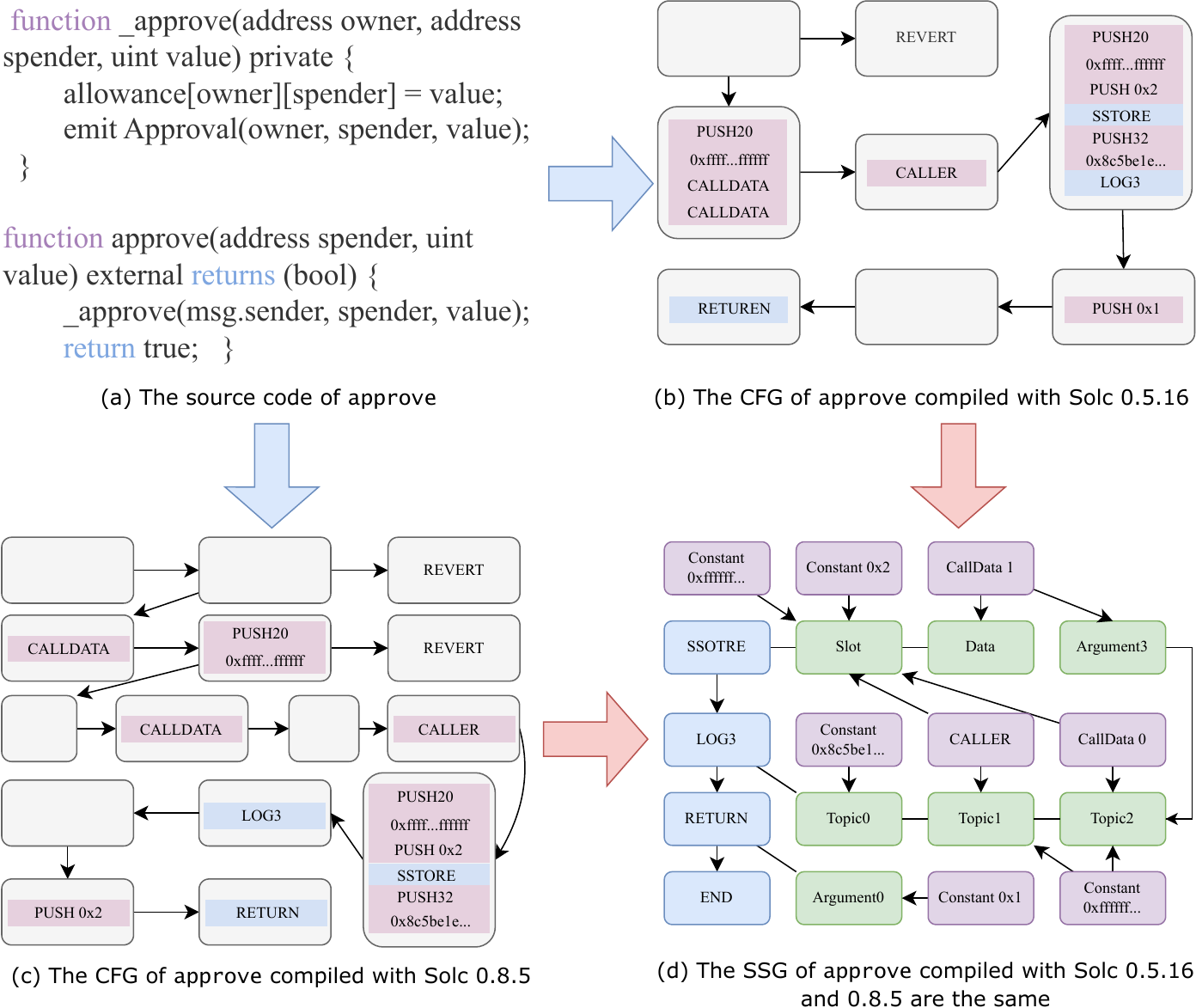}
	\caption{An example of the same source code to basic block CFG and Stable-Semantic Graph in different compiler versions.}
	\label{fig:sfgprocess}
        \vspace{-1em}
\end{figure}

\subsection{Insight of EVM Bytecode Representation}
\label{subsec:insight}

To tackle the problems, we suggest the development of an EVM bytecode representation that is capable of \textit{(i)}  effectively connecting low-level bytecode instruction with high-level semantics;
\textit{(ii)}  mitigating the impact of basic blocks reuse strategy and 
\textit{(iii)} maintaining stability across different compilation versions. During our exploration, we have two key insights: 
\begin{itemize}[leftmargin=*]
\item \textbf{Insight I: EVM instructions have graded semantic relevance.}  
EVM is a stack-based virtual machine with multiple kinds of instructions related to stack, memory, and storage. 
% According to the EVM design, all instructions will consume a certain amount of gas as a limit, which naturally leads to differences in the gas consumption of EVM instructions (\eg, storage instructions consume more gas than stack instructions). 
The EVM dictates that solely storage-related instructions are capable of altering the global state of the blockchain. Conversely, stack-related instructions are only able to modify the information within the template stack.
In DeFi ecosystem, smart contracts need to cooperate with other smart contracts (\ie, using \code{CALL}), and dApps (\ie, using \code{LOG} to send logs) to execute intricate tasks. 
This indicates that these instructions exert a more substantial influence on the blockchain, thereby having high semantic relevance. 
% We call these instructions as \textbf{stable instructions} in this paper, see details in Table~\ref{tab:keystatement}

Figure~\ref{fig:sfgprocess} shows a simple \code{approve} function from the Uniswap v2 protocol. This function increases the allowance from the owner to the spender. It uses \code{SSTORE} to update the \code{allowance} slot, emits a \code{LOG} to notify other DApps, and ends with \code{RETURN}. Evidently, these instructions are crucial for the contract to function as intended.

% For instance, we present a simple example of the \code{approve} function from the Uniswap v2 protocol in Figure~\ref{fig:sfgprocess}. The function capability is to increase the allowance from the owner to the spender. Naturally, this function utilizes \code{SSTORE} to update the \code{allowance} slot, emits a \code{LOG} to notify other DApps, and finally employs \code{RETURN} to conclude the execution. Clearly, these instructions are essential for enabling the intended functionality and ensuring the proper operation of the contract.

\item \textbf{Insight II: Instructions with high semantic relevance are more stable against the basic block reuse strategies.} 
Instructions with high semantic relevance are of greater significance and cost more than those with low semantic relevance (i.e., those related to stack or memory operations). 
This stability may arise from the significant correlation between these instructions and their associated semantics.
To ensure program correctness, Solc will reduce the reuse of basic blocks where these instructions are located, thereby maintaining stable connectivity and order. 

Figure~\ref{fig:sfgprocess} illustrates the CFG and \sfg{} of the same \code{approve} function using Solidity compilers \solc{} 0.5.16 and \solc{} 0.8.5.
The \solc{} update has resulted in a huge difference in the control flow graph of the function (Figure 2.b vs. Figure 2.c).
In contrast, the relationships between \sinst{} remain clear and concise (see Figure 2.d). 
% This stability across different compiler versions and optimization-level options inspired this study.
\end{itemize}

\begin{table}[!t]
    \caption{The stable control node.}
    \label{tab:keystatement}
    \centering
    \resizebox{.45\textwidth}{!}{
    \begin{tabular}{lc}
    \hline
    \textbf{Control Node Type}        & \textbf{Opcode}                                                                               \\ \hline
    Storage      & SSTORE, SLOAD                                                                        \\ \hline
    Log         & LOG0, LOG1, LOG2, LOG3, LOG4                                                         \\ \hline
    CALL     & \begin{tabular}[c]{@{}c@{}}CALL, STATICCALL, BALANCE \\ DELEGATECALL\end{tabular}   \\ \hline
    RETURN & \begin{tabular}[c]{@{}c@{}}RETURN, SELFDESTRUCT, \\ REVERT, THROW, STOP\end{tabular} \\ \hline
    \end{tabular}
    }
    \vspace{-1em}
\end{table}

\subsection{Formal  Definition of Stable-Semantic Graph}
\label{subsec:defination}

Based on our insights, we propose a novel EVM function bytecode representation, the \textbf{Stable-Semantic Graph} (or \sfg{} for short), to perform the similarity detection. The \sfg{} definition is proposed as follows.

\textbf{Definition I:} A \sfg{} is a directed heterogeneous graph $\mathcal{G} = (\mathcal{V},\mathcal{E})$ with a node type
mapping function $\phi$: $\mathcal{V} \rightarrow \mathcal{A}$ and a edge type mapping function $\varphi$ : $\mathcal{E} \rightarrow \mathcal{R}$.
% Specifically, the details are as follows:
\begin{enumerate}[leftmargin=*]
    \item The \sfg{} nodes $\mathcal{V}$ consists of two types of nodes: (i) Control Flow Node ($\mathcal{C}$) and (ii) Data Flow Node ($\mathcal{D}$). Consequently, the number of node types $|\mathcal{A}|$ is 2.
    \item The \sfg{} edges $\mathcal{E}$ consists of three types of edges: (i) the control flow relationship between the Control Flow Node, with the edge type $(\mathcal{C},\mathcal{C})$; (ii) the data flow relationship from the Data Source Node to Data Sink Node, with the edge type $(\mathcal{D},\mathcal{D})$; and (iii) the corresponding relationship between the Control Flow Node and Data Node, with the edge type $(\mathcal{C},\mathcal{D})$.Consequently, the number of edge relation types $|\mathcal{R}|$ is 3.
    \item $\mathcal{C}$ is a node set of stable instructions. Specifically, there are four types of instructions: \textsc{Storage Stmt}, \textsc{Log Stmt}, \textsc{Call Stmt}, and \textsc{RETURN Stmt}, see details in Table~\ref{tab:keystatement}.
    \item $\mathcal{D}$ is a node set assigned data attributes in $\Omega$, including two data types: (i) Sink Variables, which represent the values that instructions push to the stack or store in memory, that directly connect to the related instructions; and (ii) Source Variables, which represent the values that instructions pop from the stack or load from memory, that have data flow relationship with sink data nodes. The detailed Sink Variables and Source Variables are provided in Table~\ref{tab:keysource}. 
    % Path insensitivity indicates whether a data node type is a global variable, independent of the execution path. Specific variable sensitivity refers to whether a data type has a hardcode specific variable within the bytecode, see details in Section~\ref{subsubsec:dfg}. 
\end{enumerate}

% \usepackage{multirow}
% \begin{table}[]
%     \centering
%     \caption{The Key Source Variables}
%     \label{tab:keysource}
%     \begin{tabular}{|l|l|}
%     \hline
%     Variable Type& Attributes    \\ \hline
%     Constant     & Constant Number       \\ \hline
%     \begin{tabular}[c]{@{}l@{}}Environmental \\ Information\end{tabular} & \begin{tabular}[c]{@{}l@{}}Opcode in \{ADDRESS, ORIGIN, CALLER, \\ CALLVALUE, CODESIZE, GASPRICE, \\ BLOCKHASH, COINBASE, TIMESTAMP, \\ NUMBER, PREVRANDAO, GASLIMIT, \\ CHAINID, SELFBALANCE, BASEFEE\}\end{tabular} \\ \hline
%     Calldata     & Index of parameters   \\ \hline
%     \multirow{2}{*}{\begin{tabular}[c]{@{}l@{}} Definition\end{tabular}} & \begin{tabular}[c]{@{}l@{}}Opcode in \{CALL, DELEGATECALL, \\ STATICCALL\} and Index of returned values\end{tabular}  \\ \cline{2-2} 
%  & Opcode in \{SLOAD, BALANCEOF\}\\ \hline
%     \end{tabular}
    
% \end{table}

\begin{table}[]
    \centering
    \caption{The stable data node and attributes.}
    \label{tab:keysource}
\resizebox{.45\textwidth}{!}{
\begin{threeparttable}
\begin{tabular}{llccc}
\hline
\multicolumn{2}{c}{\textbf{Data Node Type}}                   & \textbf{PI}\tnote1{} & \textbf{SV}\tnote2{} & \textbf{Attributes}                      \\ \hline
\multirow{5}{*}{Source} & Constant         & \CheckmarkBold                                   & \CheckmarkBold                            & Specific Value                  \\ \cline{2-5} 
                        & Information      & \CheckmarkBold                                   & \XSolidBrush                              & Opcode                          \\ \cline{2-5} 
                        & Call data        & \CheckmarkBold                                   & \XSolidBrush                              & Offset      \\ \cline{2-5} 
                        & Return data      & \XSolidBrush                                     & \XSolidBrush                              & Offset    \\ \cline{2-5} 
                        & Definition       & \XSolidBrush                                     & \XSolidBrush                              & opcode                          \\ \hline
\multirow{4}{*}{Sink}   & Log     & \XSolidBrush                                     & -                                                        & topic index, index              \\ \cline{2-5} 
                        & Storage & \XSolidBrush                                     & -                                                        & Slot, and data for sstore       \\ \cline{2-5} 
                        & Call    & \XSolidBrush                                     & -                                                        & address, value, selector, index \\ \cline{2-5} 
                        & Return  & \XSolidBrush                                     & -                                                        & index                           \\ \hline
\end{tabular}
\begin{tablenotes}
\small
\item[1] Path insensitivity (PI) indicates whether a data node type is a global variable, independent of the execution path. 
\item[2] Specific variable (SV) sensitivity refers to whether a data type has a hardcode specific variable within the bytecode
\end{tablenotes}
\end{threeparttable}
}
\vspace{-1em}
\end{table}
\section{Esim Design Overview}
\label{sec:designoverview}

In this section, we present an overview of the design of our EVM bytecode similarity detection system, which we call \system{} in this paper.
Specifically, \system{} first statically builds precise \sfg{} for each function in the contract, and then applies a neural-network-based graph-embedding method to represent \sfg{}.
% Figure~\ref{fig:overview} shows the architecture of \system{}.

The \system{} consists of two modules, \ie, 
\textit{\sfg{} Builder} and \textit{\sfg{} Embedding Generator}, as follows:
\begin{itemize}[leftmargin=*]
\item \textbf{\sfg{} Builder.} This module accepts EVM contract bytecode as input and outputs the corresponding \sfg{}s.
It consists of two sub-processes: \textit{stable control flow graph  (\kcfg{}) construction} and \textit{stable data flow graph (\kdfg{}) integration}. Specifically, the first sub-process parses each function from the contract bytecode, identifies all \sinst{}s, and determines their control flow relationships. Then, the second sub-process identifies all sink data nodes within the stable instructions and incorporates data flow information through taint analysis. 
\item \textbf{\sfg{} Embedding Generator.} Consider the \sfg{} as a heterogeneous graph, which consists of two kinds of nodes, and three kinds of edges. This module leverages the heterogeneous graph network to train an embedding model. The embedding result is used to represent the \sfg{}.
\end{itemize}
Finally, the EVM bytecode similarity can be quantitatively measured by the similarity of the vector expression of \sfg{}. In the following, we detail these two modules in Section~\ref{subsec:stg} and Section~\ref{subsec:embed}, respectively.

% \begin{figure}[!t]
% \centering
% \includegraphics[width=.45\textwidth]{Fig/ESIM-overview-narrow.drawio.pdf}
% 	\caption{The overview design of Esim.}
% 	\label{fig:overview}
%         \vspace{-1em}
% \end{figure} 

\subsection{SSG Builder}
\label{subsec:stg}

In this section, we introduce the design of the \sfg{} builder. Specifically, the \sfg{} builder has two sub-processes: \textit{(i)} stable control flow graph (\kcfg{}) construction, and \textit{(ii)} stable data flow graph (\kdfg{}) integration.
First, in the \kcfg{} construction sub-process, the contract bytecode is taken as input, and split into several external functions. This sub-process then outputs the \kcfg{} for each external function. 
Second, the \kdfg{} integration sub-process takes the previously generated \kcfg{} as input and enhances it by attaching the variable and data flow information, thereby adding semantic richness. Finally, the \kdfg{} interaction outputs the \sfg{} for each function.

\subsubsection{Stable Control Flow Graph Construction}
\label{subsubsec:dfg}
As mentioned in Section~\ref{sec:background}, the EVM bytecode mixes all functions without obvious function boundaries, and the relationship of basic blocks is unclear. Specifically, to achieve \kcfg{} construction, there are three phases, as follows:
\begin{itemize}[leftmargin=*]
    \item \textbf{Locating function entrypoints.}
The EVM functions can be categorized into two types: \textit{external} and \textit{internal} functions.
\textit{(i) External functions} serve as interfaces for the usage of external entities usage, such as Externally Owned Accounts (EOAs) and other smart contracts. 
% To invoke a specific external function, users should use a designated function selector~\footnote{It is the first (left, high-order in big-endian) four bytes of the Keccak-256 hash of the signature of the function\cite{solidity_selector}.} along with the corresponding calldata.
EVM executor uses the function selector~\footnote{It is the first (left, high-order in big-endian) four bytes of the Keccak-256 hash of the signature of the function\cite{solidity_selector}.} to locate the correct function call offset, ensuring a precise execution.
\textit{(ii) Internal functions} are accessible only within the contract itself. According to the \solc{} compilation strategy, internal functions are fully inlined into external functions, effectively becoming part of them and making them inseparable. 
Therefore, in this study, we only focus on external functions, while internal functions are already inlined.

The \solc{} compiler utilizes fixed paradigms~\cite{contro2021ethersolve}, matching calldata with external function selectors to determine the entrypoint of a function. 
% As such, by following these paradigms, we can map the correspondence between function selectors and their respective entry-point offsets for functions.
In the algorithm~\ref{alg:kscfg}, we use \texttt{getFunctions(bytecode)} to represent the method to obtain all functions from a contract.

\item\textbf{Building function-level CFGs.}
% After identifying the functions of each function, the next step is to construct the CFG for these functions.
% This involves two sub-tasks: first, building the global CFG, and then partitioning it into function-level CFGs to delineate the boundaries of individual functions.
As the EVM operates as a stack-based machine, where the control flow is determined by the values stored in the stack before executing \textit{jump} instructions (include \code{jump} and \code{jumpi}), the SOTA EVM decompilation technique~\cite{grechGigahorseThoroughDeclarative2019} leverages simulated execution to identify the stack variables that correspond to jump destinations, facilitating the construction of the CFG. 
To determine the boundaries of each function, previous studies~\cite{grechGigahorseThoroughDeclarative2019,grechElipmocAdvancedDecompilation2022,contro2021ethersolve,brent2018vandal} use methods such as simulated execution and constraint solving to verify whether a basic block is genuinely reachable within a function by exploring various execution paths.

% The basic block CFG construction is a fundamental problem in the EVM static analysis area. There have been several studies~\cite{grechGigahorseThoroughDeclarative2019,grechElipmocAdvancedDecompilation2022,contro2021ethersolve,brent2018vandal} to perform EVM bytecode static analysis. 

For our system, we adopt the SOTA framework, Elipmoc~\cite{grechElipmocAdvancedDecompilation2022}, to construct function-level CFGs. In this study, we further enhance clarity and precision by cloning and inlining reused basic blocks separately for each function. 
In Algorithm~\ref{alg:kscfg}, \texttt{getCFG(bytecode, function)} represents the method to generate the CFG of a specific function, while \texttt{getPredecessors(block,cfg)} is used to retrieve the predecessors of a given basic block within the CFG.

\item\textbf{Extracting the \kcfg{} from the function-level CFG.} 
% After the above two phases are completed, we obtain the basic block CFG of the EVM bytecode. 
Focusing on each basic block, we extract all the stable instructions within these blocks, denoted as \texttt{getStableStmts(block)} in Algorithm~\ref{alg:kscfg}. For instructions residing within the same block, we establish control flow connections according to their sequential order in the code. For instructions located in different blocks, we iteratively identify the last reachable block, thereby facilitating the appropriate control flow connections, which is denoted as \texttt{getPredecessors(block,cfg)} in Algorithm~\ref{alg:kscfg}.
\end{itemize}

In the end, the Algorithm~\ref{alg:kscfg} outlines the process of constructing the \kcfg{}s for a given contract bytecode. 
% It takes the contract bytecode as input and returns a mapping of functions to their respective \kcfg{}s, represented as a tuple comprising nodes and edges. The construction of the \kcfg{} is accomplished by traversing the CFG and transforming the relationships between blocks into relationships between stable instructions.

\begin{algorithm}[!t]
    \footnotesize
	\SetAlgoLined
	\SetKwData{True}{true}
	\KwData{$bytecode$: contract bytecode.}
	\KwResult{$\kcfg{}=\{function \rightarrow (Nodes,Edges)\}$: the stable control flow graph of $bytecode$.}
	
    \SetKwFunction{Fprestmt}{resolvePreStableStmts}
    \SetKwFunction{Fgetkeystmt}{getStableStmts}
    \SetKwFunction{Fgetpre}{getPredecessors}
    \SetKwFunction{Fgetblocks}{getBlocks}
    \SetKwFunction{Fgetfuncs}{getFunctions}
    \SetKwFunction{Fgetcfg}{getCFG}

    \SetKwFunction{Fckscfg}{construct\kcfg{}}
    \SetKwProg{Fn}{Function}{:}{}
    
    \Fn{$\Fprestmt{block, cfg, visited}$} {
        $predStmts \gets \{\}$  \Comment{The predecessor stable stmts of the block} \newline 
        $predBlocks \gets \Fgetpre{block, cfg}$ \Comment{Get all predecessor blocks}  \newline
        \For{$pb$ in $predBlocks$}  { 
            $keyStmts \gets \Fgetkeystmt(pb)$ \Comment{Get all stable stmts in the block} \newline
            \uIf{$keyStmts \ne \emptyset$} { 
                $predStmts \gets predStmts \cup \{\Fgetkeystmt(pb)[-1]\}$ 
            } \uElseIf {$pb$ $not$ $in$ $Visited$} {
                $visited \gets visited \cup \{pb\}$
                \newline
                $predStmts \gets predStmts \cup \Fprestmt(pb, cfg, visited)$ 
            }
        }
        \KwRet $predStmts$
    }

    \Fn{$\Fckscfg{bytecode}$} {
        $\kcfg{} \gets Mapping()$ \Comment{The mapping of function to \kcfg{}} \newline
        \For {$function$ in $\Fgetfuncs(bytecode)$}{
            \For{$cfg$ in $\Fgetcfg(bytecode, function)$ } {
                $Nodes \gets \{\}$ \Comment{The \kcfg{} nodes} \newline 
                $Edges \gets \{\}$ \Comment{The \kcfg{} edges} \newline
                \For{$block$ in $\Fgetblocks{cfg}$}{
                    $prevStmts \gets \Fprestmt{block, cfg, \{\}}$ \newline 
                    \For{$stmt$ in $\Fgetkeystmt{block}$}{
                        $Edges \gets Edges \cup prevStmts \times \{stmt\}$ \Comment{Connect the stable control flow nodes}  \newline
                        $prevStmts \gets \{ stmt \}$ 
                    } 
                }
                $\kcfg{}[function] \gets (Nodes, Edges)$ \Comment{Update the \kcfg{}} 
            }
        }
        \KwRet $\kcfg{}$
    }
    
    \caption{The Stable Control Flow Graph Construction Algorithm}
  \label{alg:kscfg}
\end{algorithm}

\subsubsection{Stable Data Flow Graph Integration}
\label{subsubsec:kdfgbuild}

After getting the \kcfg{}, we then integrate \kdfg{} information to enhance the semantics. The main components are divided into the following three phases:

\begin{itemize}[leftmargin=*]
    \item \textbf{Locating sink data node.} The bytecode semantics are closely tied to the operators associated with stable instruction nodes in the \kcfg{}, such as the storage slot in \code{SSLOAD} and the topics content in \code{LOG}. 
    To account for the impact of these critical variables, we identify all data variables utilized in the stable instructions, and define these nodes as \textbf{sink data nodes}.
    Based on the EVM opcode document~\cite{EVMcode}, we enumerate all sink data nodes and their corresponding instruction nodes in Table~\ref{tab:keysource}. 
    % In the algorithm~\ref{alg:kvdfg}, we use \texttt{getSinks(instruction)} to represent the method to obtain all sink variables.

    \item \textbf{Backward taint analysis.}
    Only identifying the sink data nodes is insufficient. Typically, the actual content of these sink variables may originate from other variables. For example, the actual \code{selector}, \code{arguments}, and \code{return values} of a called function are typically transferred through memory from the user calldata. Therefore, it is crucial to trace the origins and propagation of these variables. 
    
    In this phase, we identify all variable nodes that may influence the sink data node and define them as \textbf{source data nodes}. 
    Depending on their different attributes, we categorize the source data nodes into four categories: \textit{Constants}, \textit{Environment Variables}, \textit{Calldata}, and \textit{Definition Variables}. 
    Among these four categories, \textit{Definition Variables} are location-sensitive, meaning the value of the same variable may be affected by its position; for instance, the return value of two identical \code{call} operations may differ due to changes in the on-chain environment. Conversely, the other categories are location-insensitive, as the values of these variables remain constant within the same transaction and cannot be altered.
    % Besides, only \textit{Constants} are defined directly using the \code{push} instruction, with their values hard-coded in the bytecode. In contrast, the other categories only possess type information and do not have specific values. In our study, we record the values of \textit{Constants}. To avoid the influence of the solidity compiler optimization, we also introduce constant folding, where only the final results of constant folding are treated as a source variable, \eg, the \code{0xfff AND 0x4} is treated as \code{0x4}. 
    
    Subsequently, we perform backward taint analysis starting from the sink data nodes to identify the relationships between the source data nodes. The backward taint analysis algorithm draws upon traditional static analysis research~\cite{dataflowanalysis}, modeling memory and considering the process of value propagation through memory~\cite{lagouvardos2020precise}. 
    % For a specific sink data node, we use the \texttt{getSources(sink)} function in Algorithm \ref{alg:kvdfg} to represent the source data nodes that are accessible to the sink data node.
    \item \textbf{Integrating \kdfg{} and building the \sfg{}.}
    The input for the \kdfg{} integration algorithm comprises the contract bytecode and the specific \kcfg{}. Initially, set \sfg{} to empty. Then, traverse the instruction nodes of each \kcfg{} and identify any contained sink data nodes. If a sink data node is found, we then locate all reachable source data nodes and record the data flow relationships. Finally, integrate the local \kdfg{} into the \sfg{} based on the declaration relationships between the sink nodes and the instruction nodes. 
    The output is the \sfg{} of the specific function.
\end{itemize}

\subsection{\sfg{} Embedding Generator}
\label{subsec:embed}

We first introduce the code similarity embedding problem in Section~\ref{subsubsec:codeproblem} and then present our heterogeneous graph embedding model in Section~\ref{subsubsec:overview}.

\subsubsection{Code Similarity Embedding Problem}
\label{subsubsec:codeproblem}

Inspired by previous studies~\cite{xu2017neural,he2020characterizing}, the code similarity embedding problem can be formed as an independent task: there exists an oracle $\pi$ determining the code similarity metric for a given task. Given two binary program functions $f_1$, $f_2$, $\pi(f_1, f_2) = 1$ indicates that they are similar; otherwise, $\pi(f_1, f_2)$ = -1 indicates that they are dissimilar.
In general, $\pi$ is determined from the intuitive judgment of domain experts and has no fixed rules and basis. 
The binary code embedding problem can be refined to find a mapping $\phi$, which maps a function $f$ to a vector representation $\mu$. Such a mapping should be consistent with the intuition $\pi$ of the domain experts. That is, given an easy-to-compute similarity function $Sim()$, (\eg, cosine distance), and two binary functions $f_1$, $f_2$, $Sim(\phi(f_1)$,$\phi(f_2))$ is close to 1 if $\pi(f_1, f_2)$ = 1, and is close to -1 otherwise. 

In our study, we assume that the binary code of a function $f$ is represented by its \sfg{}. And the embedding problem can be defined as finding a mapping $\phi$ that embeds the \sfg{} into a vector representation $\mu$. The similarity calculation between the resulting vectors $\mu$ should align with the intuitive judgments of domain experts. 
Furthermore, since the vector representation $\mu$ contains sufficient information, it can be effectively utilized for downstream tasks. 

\subsubsection{Heterogeneous Graph Embedding Model}
\label{subsubsec:overview}

As mentioned above, we presented the \sfg{} as a direct heterogeneous graph $\mathcal{G}$, combined with control flow and data flow information in $\Omega$. In this section, we explain our design of the overall network architecture to train a graph embedding for similarity detection.

\noindent\textbf{Attributes Encoding}
We employ different encoding methods to handle attributes of varying dimensions before applying the graph embedding network.
In detail, our \sfg{} attributes $\Omega$ can be classified into two types: \textit{(i)} finite field properties (\eg, opcode, data type) and \textit{(ii)} non-finite field data (\eg, immediate number, hard-code string). 
For finite field properties, we utilize one-hot encoding~\cite{rodriguez2018beyond}, a well-adopted encoding method. 
For non-finite field data, such as hard-coded strings, we utilize EVM's 256-bit binary limit to partition continuous binary data into multiple 256-bit segments.
We format all non-finite field data into binary encoding, and then crop or pad them to 256-bit format. This ensures the proper handling of binary-encoded data within non-finite fields.

\noindent\textbf{Network Learning Architecture}
After the attributes encoding, considering the mixed information of control flow and data flow, we adopt the previous heterogeneous graph embedding studies~\cite{han2019,hu2020heterogeneous,fu2020magnn} to better retain information at different latitudes. 
Given the extensive research in the field of heterogeneous graph embedding, it is beneficial for us to represent \sfg{} as a feature vector that captures and preserves a significant amount of information.

In particular, we use the Siamese architecture~\cite{chen2021exploring} combined with the graph embedding network. 
The graph embedding network takes \sfg{} $g$ as its input and outputs the embedding $\phi(g)$. 
The Siamese architecture uses two identical graph embedding networks at the top. It treats the \sfg{} embedding $\phi(g_i)$ ($i$ = 1, 2) as input and outputs the embedding distance.
Notably, the embedding networks share the same set of parameters; thus, during training, the two networks remain identical.
Given a set of $\mathcal{K}$ pairs of \sfg{} $<g_i$, $g^{'}_i>$, with ground truth label information $y_i \in \{-1,+1\}$, where $y_i$ = +1 indicates that $g_i$ is similar to $g^{'}_i$, and $y_i$ = -1 otherwise. The Siamese network output for each pair is defined as:

\begin{equation}
\begin{aligned}
    Sim(g,g^{'}) = distance(\phi(g), \phi(g^{'}))
\end{aligned}
\end{equation}

where $\phi$ is the embedding network model, which converts the \sfg{} to vectors. 
Then, for all the $\mathcal{K}$ pairs of \sfg{}, we use the following loss function to train our embedding network. Specifically, we $\varepsilon^+$ represent the similar pairs set, where the $(u,v) \in \varepsilon^+, y(u,v) = 1$, and the $\varepsilon^-$ represent the unsimilar pairs set, where the $(u^{'},v^{'}) \in \varepsilon^-, y(u^{'},v^{'}) = -1$

\begin{equation}
\begin{aligned}
    \mathcal{L} = - ( \frac{1}{|\varepsilon^+|} \sum_{(u,v) \in \varepsilon^+} Sim(g^u,g^v)  \\ + \frac{1}{|\varepsilon^-|} \sum_{(u^{'},v^{'}) \in \varepsilon^-} (1-Sim(g^{u^{'}}, g^{v^{'}}))
    )
\end{aligned}
\end{equation}

The goal of training is to minimize the loss of the function. Upon completion of the training, we freeze the parameters and employ the embedding function $\phi(g)$ as the final model.

\section{Evaluation}
\label{sec:evaluation}

In this study, we have implemented a prototype of \system{}, and further evaluate \system{} by answering the following research questions: 
\begin{itemize} [leftmargin=*]
 \item  \textbf{RQ1}: What is the accuracy of \system{} in \sfg{} construction. 
 \item  \textbf{RQ2}: What is the effectiveness of \system{} in EVM bytecode similarity detection.  
 \item  \textbf{RQ3}: What is the efficiency of \system{} on large-scale data.
 \end{itemize}
% \lei{Not questions. Using what, does...}
To facilitate these evaluations, we have curated several datasets. Across all evaluations, our approach demonstrates significant advantages over state-of-the-art (SOTA) methods.
\subsection{Evaluation Setup}
\label{subsec:eva_setup}

We build our \sfg{} extractor based on Gigahorse, an open-source EVM bytecode static analysis tool. And we implement the neural network model in Pytorch in Python.
Our experiments are conducted on a server equipped with two Intel Xeon Silver 4214R CPUs running at 2.4GHz, 377 GB memory, 3TB SSD, and 1 NVIDIA GeForce RTX 3080 GPU.

\noindent \textbf{Datasets.}
In our evaluation, we collect two datasets: 
\textit{(i)} Dataset I for evaluating the accuracy of \sfg{} generation;
\textit{(ii)} Dataset II for training the neural network and evaluating the accuracy of the \system{}; 
\textit{(iii)} Dataset III for evaluating the real-world performance \& efficiency on large-scale dataset of the \system{}.

\begin{itemize}[leftmargin=*]

    \item \textbf{Dataset I.}   
    To evaluate the accuracy of \sfg{} generation, we construct a dataset with two well-known open-source projects in Ethereum, \ie, the Uniswap v2 contract, and the USDT. Our dataset includes both the Solidity source code and corresponding binary representations. 
    
    \item \textbf{Dataset II.}  This dataset is used for neural network training. To ensure we have the ground truth, this dataset consists of bytecode compiled from public source code.
    This approach is commonly employed in tasks related to binary similarity, that is, we consider two \sfg{} compiled from the same source code function to be similar, and those from different functions to be dissimilar.

    % In addressing EVM bytecode, we highlight the challenge associated with this aspect of our work. The \solc{} compiler has undergone significant changes, and even with access to the source code, compiling across different versions and compilation options can still result in numerous errors.
    % To tackle this challenge, we have proposed an approach for generating the dataset.
    % Specifically, we conducted a manual review of the syntax changes introduced by various versions of \solc{}. We then corrected these syntax errors using the \solc{} Abstract Syntax Tree (AST) and regular expressions. Ultimately, we modified the source code to ensure compatibility with each \solc{} version. 
    To ensure coverage of our dataset, we selected the final sub-version of each \solc{} version, \ie, \code{0.5.17}, \code{0.6.12}, \code{0.7.6}, and \code{0.8.20} versions. We also configured different optimization options such as \code{none}, \code{opt=200},  \code{opt=200000} and \code{via-ir} introduced in version \code{0.8}. Notably, \solc{} versions before 0.5 are excluded from our datasets because they are outdated and not relevant to modern DeFi protocols. 
    The focus on more recent compiler versions is supported by the timeline of the emergence of DeFi, particularly with DeFi Summer in 2020~\cite{Defisummer}. 
    % For instance, Uniswap V2, one of the earliest and most prominent DeFi protocols, uses \solc{} 0.5.
    
    We selected the source code from the top 1,000 Ethereum DeFi projects~\footnote{Given the substantial popularity of these DeFi projects, 907 of these projects are open source.} ranked by DefiLlama~\cite{Defillma}.
    In total, we obtain 21,018 \sfg{}s~\footnote{Converting from higher versions to lower versions can lead to unresolved errors. To preserve the original syntax's meaning, we consciously decided not to make additional modifications that could alter the semantics.}. We split the Dataset into three disjoint subsets of functions for training (70\%), validation (20\%), and testing (10\%), respectively. 
    % By doing so, we guarantee that one function compiled from the source code will be used only once and the pre-trained model can generalize to unseen functions.

    \begin{table}[tb]
\centering
\caption{The number of \sfg{} in Dataset II}
\label{table:traindataset}
\resizebox{.45\textwidth}{!}{
\begin{tabular}{c|l|c|c|c|c}
\hline
Solc Version & 0.5.17  & 0.6.12  & 0.7.6  & 0.8.20   & 0.8.20 (viair) \\ \hline
None      & 650  & 923  & 1241 & 3877  & -          \\ \hline
Opt\_200    & 591  & 890  & 1097 & 3368  & 877        \\ \hline
Opt\_200000 & 634  & 930  & 1157 & 3863  & 920        \\ \hline
Total     & 1875 & 2743 & 3495 & 11108 & 1797       \\ \hline
\end{tabular}%
}
\vspace{-1em}
\end{table}

    \item \textbf{Dataset III.} 
    This dataset is used for large-scale real-world evaluation. It encompasses a large-scale collection of contract bytecode that were deployed on Ethereum, Optimisim, Binance Smart Chain (BSC), Polygon POS, Arbitrum One, and Avalanche-C-Chain between January 1st, 2023 and June 13th, 2024.
    In total, we collect 2,675,573 distinct contracts and their corresponding bytecode.
    
\end{itemize}
\noindent \textbf{Baselines} We select the following baselines: 
\begin{itemize}[leftmargin=*]
    \item \textit{FuzzHash}~\cite{he2020characterizing}, a similarity detection method that relies on a structural-aware hash technique to compare the similarity between two function bytecodes.
    \item \textit{Genius}~\cite{feng2016scalable} propose the ACFG to represent the function bytecode. As the original Genius was initially designed for C++, we have re-implemented Genius for EVM bytecode. Specifically, we utilize the basic block CFG and incorporate node attributes to construct the Attribute Control Flow Graph (ACFG) (due to page limit, see more details in our repository). We apply the representation and train a corresponding model using \textbf{Dataset II}.
    \item \textit{Eclone}~\cite{liu2018eclone}, a similarity detection system based on small code snippets, supports only 0.4 \solc{} versions and has not been updated in six years. 
    As a result, it is incompatible with the current EVM environment and our dataset, limiting its utility to compare modern DeFi contracts. Thus, we reference Eclone's results in this paper solely for context. 
    Importantly, this comparison is conservative, since Eclone relies on small symbolic sketches, which restricts its ability to capture the high-level semantics required for complex modern DeFi contracts. 
    % Its performance, while adequate for simpler legacy contracts, would deteriorate significantly when applied to contemporary, more sophisticated DeFi contracts.
    \item \textit{Etherscan}~\cite{Etherscan} is the most popular industry EVM similarity detection tool. Etherscan only compares bytecode similarity on a contract granularity, so we only compare it on our real-world dataset. 
\end{itemize}

\begin{figure*}[!tb]
  \centering
  \subfigure[Loss vs. Epoches]{\includegraphics[width=0.24\textwidth]{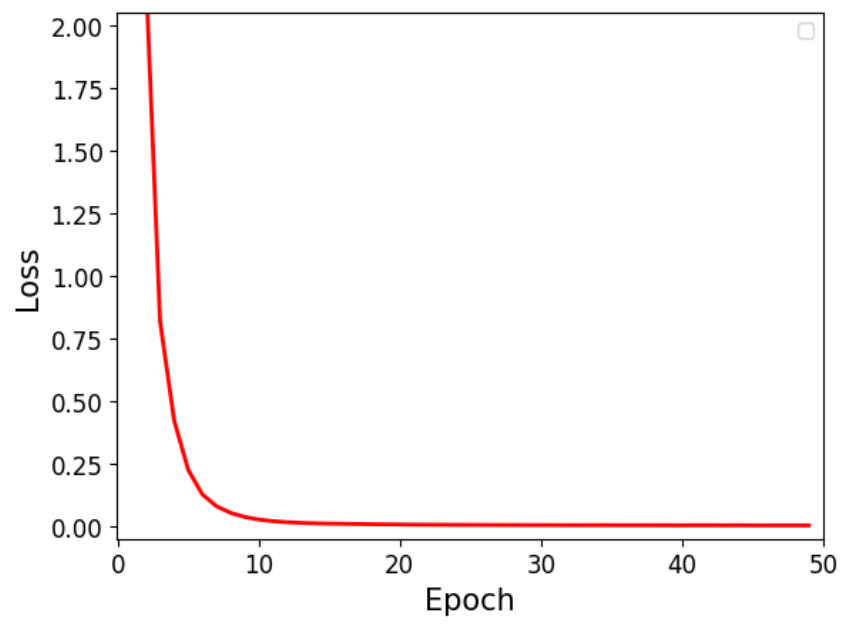}}
  % \hfill
  \subfigure[AUC vs. Embedding Layers]{\includegraphics[width=0.24\textwidth]{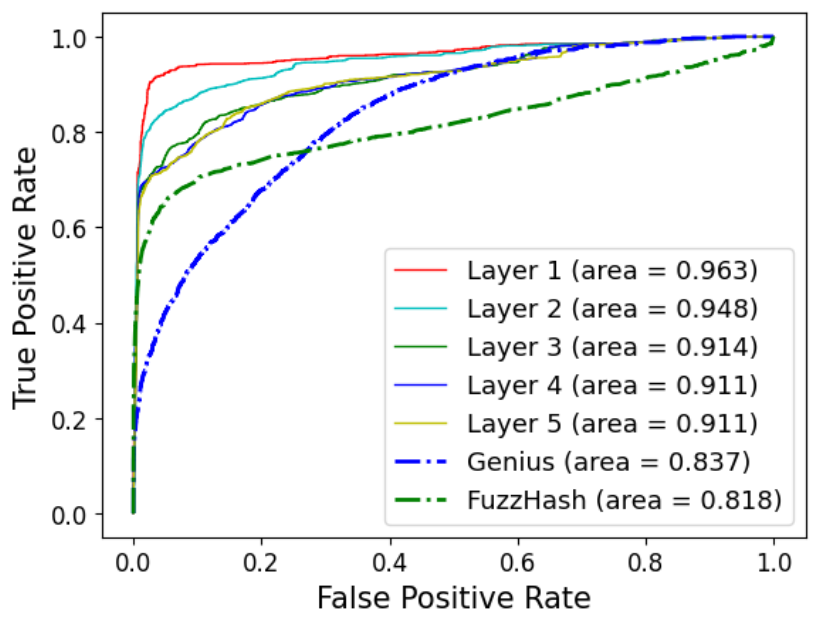}}
  % \hfill
  \subfigure[AUC vs. Embedding Size]{\includegraphics[width=0.24\textwidth]{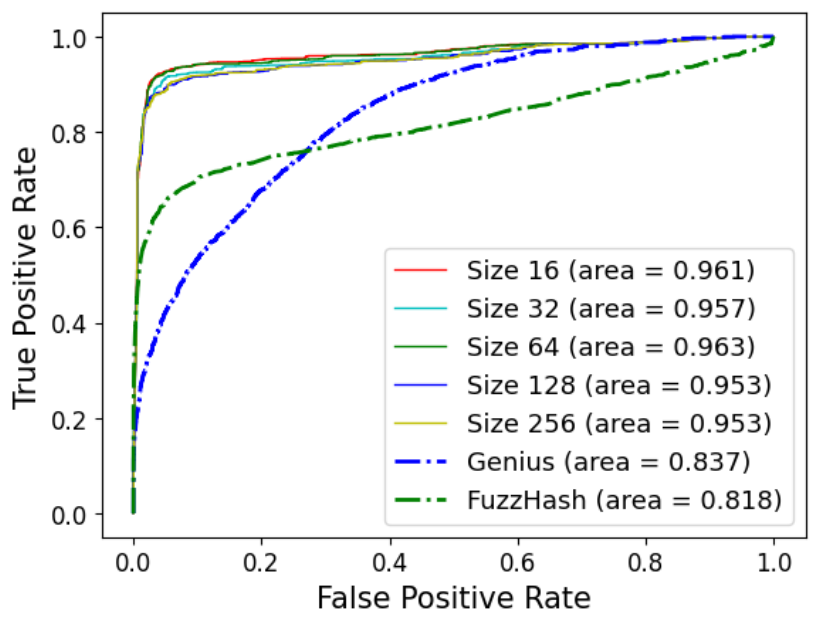}}
  % \hfill
  \subfigure[AUC vs. SSG Components]{\includegraphics[width=0.24\textwidth]{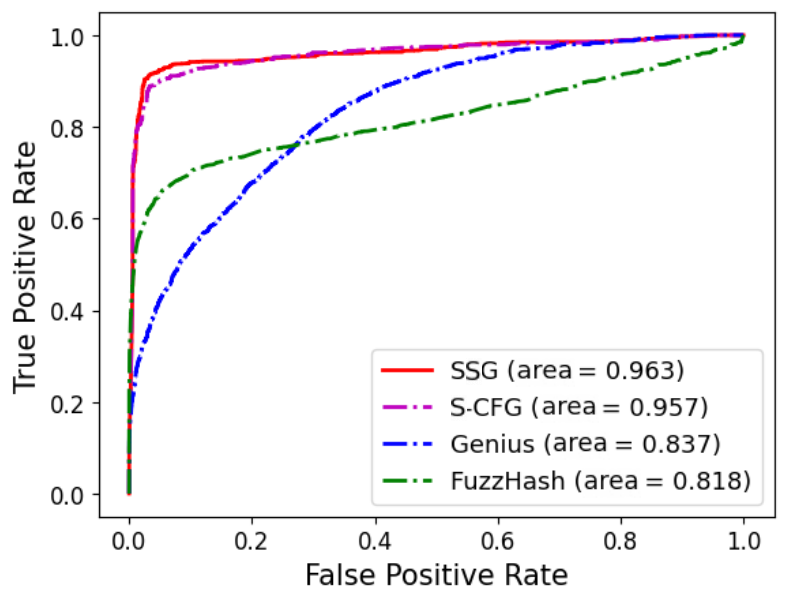}}
  \caption{The effectiveness of different hyperparameters of heterogeneous network.}
  \label{fig:subfigures}
  \vspace{-1em}
\end{figure*}

\subsection{\sfg{} Generation Accuracy}
\label{subsec:accuracy}

First, we evaluate the accuracy of the generation of the \sfg{}. To this task, we evaluate the accuracy of \sfg{} generation using \textbf{Dataset I}. 
In particular, we measure the accuracy for the following two tasks: \textit{(i)} the accuracy of the stable instruction node's control flow graph; \textit{(ii)} the accuracy of the stable instruction node's data flow graph.
Note that, due to the absence of a ground-truth evaluation dataset, we have implemented a detailed methodology to ensure the reliability of our evaluation experiments.

Firstly, the contracts in the dataset include both source code and bytecode files, enabling our engineers to analyze the correspondence between the source code and bytecode instructions. The availability of source code provides a more reliable and intuitive foundation for human understanding and interpretation.
We then manually analyze the instructions within each function, cross-referencing the corresponding source code to determine the \kcfg{} and \kdfg{}. During this process, three engineers~\footnote{All engineers have more than three years of experience in blockchain security and possess extensive expertise in code security analysis.} work independently to ensure thorough and unbiased analysis.
Finally, according to the majority principle, our engineers collaborate to reach a consensus, establishing the manually reviewed ground-truth results.

% Table~\ref{tab:graph_eva} presents the evaluation results for \sfg{} construction. 
To evaluate \system{}, we randomly selected five functions from each contract. After manual review, the control flow construction achieved an F1-score of 100\%. For data flow identification, \system{} identified 590 true positive (TP) edges, 51 false positive (FP) edges, and 9 false negative (FN) edges. Notably, even the lowest F1-score for individual functions was 89.4\%, while the overall F1-score reached 95.16\%. 
% Our \sfg{} is a combination of \kcfg{} and \kdfg{}, and the high performance in identifying each edge underscores the accuracy of \system{} in the construction of the \sfg{}.

\begin{tcolorbox}[size=title,colback=white]
\textbf{Answer to RQ1:} \system{} achieves high accuracy in \sfg{} construction, with 100\% F1-score in \kcfg{} construction and 95.16\% F1-score in \kdfg{} construction.
\end{tcolorbox}

\subsection{Graph Embedding Model Evaluation}
\label{subsec:modelevaluate}
In this section, we evaluate the graph embedding model. Firstly, we provide details of the training parameters. Secondly, we assess the accuracy of our model and compare it with baseline studies. Thirdly, we investigate the effect of the hyperparameters on the performance of our model. Finally, we conduct a visualization experiment to demonstrate the effectiveness of our approach.

\subsubsection{Training Details} Our heterogeneous network model is first pre-trained using \textbf{Dataset II} training dataset. We use the Adam optimization algorithm and set the learning rate to 0.001. We use the Siamese architecture, and each mini-batch contains 100 pairs. To ensure our training coverage, we iterate over all training sets in one epoch, and the mini-batch training data is randomly shuffled before being fed to the training process.
After every epoch, we measure the loss and Area Under Curve (AUC) on the validation set. By default, the heterogeneous model is a heterogeneous GNN, the training epochs are 50, the embedding size $p$ is 64, and the depth $n$ is 1.

\subsubsection{Accuracy}
We evaluate the accuracy of the \system{}. To this end, we use the testing dataset in \textbf{Dataset II} with the ground truth labels (\ie, from the same source function vs. not). This similarity-testing dataset consists of 14,286 pairs. Figure~\ref{fig:subfigures} illustrates the AUC for \system{} and two baseline approaches. The AUC of \system{} is 0.963 and we can see that \system{} outperforms both \textit{FuzzHash} (0.818) and \textit{Genius} (0.837) by a large margin. 

Turning to \textit{Eclone}, it is not open source and has limitations in terms of compatibility, as it only supports \solc{} versions below 0.4. To provide a comparison, we directly utilized the AUC value (0.94) obtained on its own dataset from the original paper. 
It is worth mentioning that its dataset is dichotomous, comprising only optimized or unoptimized functions for comparison. In contrast, our dataset is more comprehensive and extensive, encompassing \solc{} versions ranging from 0.4 to 0.8 and incorporating three different optimization levels.
Despite the complexity and scale of our dataset, our system outperforms \textit{Eclone} with a higher AUC value (0.963 versus 0.94). This result demonstrates the superior effectiveness of our system.

\subsubsection{Hyperparameters}
In this subsection, we evaluate the impact of hyperparameters in our model. In particular, we examine the impact of the number of training epochs, embedding depth, embedding size, and \sfg{} component.
\begin{itemize}[leftmargin=*]
    \item \textbf{Number of epochs.} 
    We train the model for 50 epochs and evaluate the model over the validation set every epoch for the loss. The results are plotted in Figure~\ref{fig:subfigures}. From the figures, we can observe that the loss drops to a low value after 10 training epochs, and then almost remains the same. The lowest loss (0.0034) appears after the model is trained for 40 epochs. 
    \item \textbf{Number of embedding layers.}
    We vary the number of layers in the heterogeneous GNN model. From Figure~\ref{fig:subfigures}, we observe that when the embedding depth is 1, the AUC has the highest value. And we can also observe that adding more layers does not increase the AUC.
    \item \textbf{Embedding size.}
    We evaluate the effect of increasing the size of the model's embedding. From the result, we find that the best AUC is achieved at 256 embedding size. However, all curves corresponding to embedding sizes no smaller than 64 are close to each other. Since a large embedding size requires longer training time, choosing the embedding size to be 64 is a good trade-off between performance and efficiency.
    \item \textbf{\sfg{} component.}
    To evaluate the \sfg{} component effectiveness, we separate the \sfg{} into \textit{(i)} the \kcfg{} only (K-CFG), and \textit{(ii)} the \sfg{}. Based on the findings presented in Figure~\ref{fig:subfigures}, it is evident that \sfg{} outperforms other approaches. This success can be attributed to the integration of both control flow and data flow information, which enhances the overall performance of the model.
    
\end{itemize}

\subsubsection{Visualization}
\label{subsec:visual}
We visualize the graph embedding model to understand its effectiveness. Specifically, we select five well-known source functions from decentralized finance protocols: \code{burn}, \code{flash}, and \code{swap} from Uniswap v3; \code{redeem} and \code{borrow} from Compound v2. We then compile the source code of these functions using various compiler versions and optimization levels to assess differences.
We then use t-SNE~\cite{van2008visualizing} to project the high-dimensional embeddings onto 2D embeddings. We plot the projected points in our repository (due to page limit) with different source functions in different colors.
We can observe that (1) bytecode functions compiled from the same source code are close to each other, indicating that our embeddings perform well for cross-version and cross-optimization problems; and (2) bytecode functions compiled from different source functions are far from each other, indicating that our embeddings are semantic-aware and effective. Therefore, this visualization illustrates that our embedding model can preserve the semantic information of the function.

\begin{tcolorbox}[size=title,colback=white]
\textbf{Answer to RQ2:} \system{} is effective in EVM bytecode similarity detection with 0.963 AUC (at least 0.126 higher than previous studies). Meanwhile, \system{} is insensitive to hyperparameters (only 0.028 effectiveness of AUC).
\end{tcolorbox}

\begin{figure}[!t]
  \centering
  \subfigure[\sfg{} generation\&embedding time]{
    \includegraphics[width=0.21\textwidth]{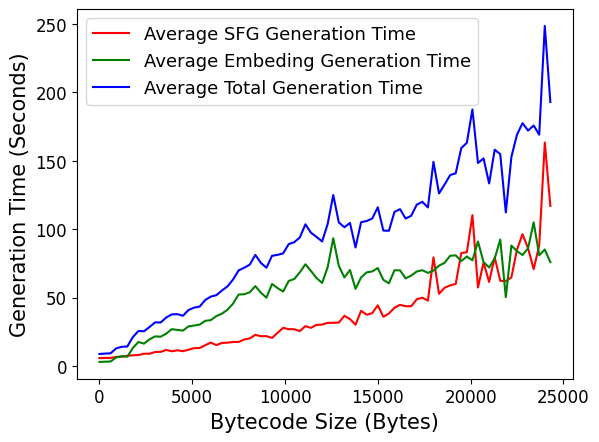}
  }
  % \hfill
  \subfigure[Database query time]{
    \includegraphics[width=0.21\textwidth]{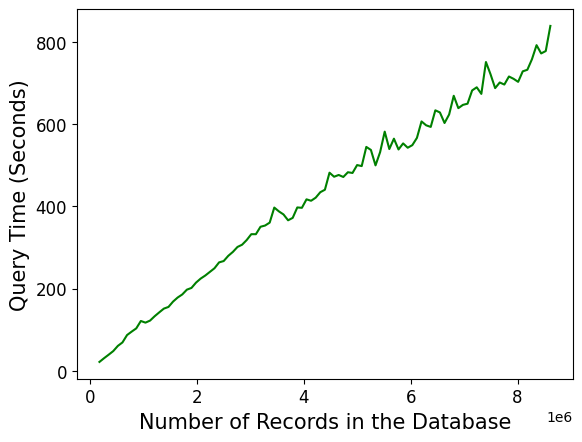}
  }
  
  \caption{The time efficiency evaluation of \system{}, including the \sfg{} generation, embedding time and the database query time.}
  \label{fig:dotvecdbtime}
  \vspace{-1em}
\end{figure}

\subsection{Efficiency}
\label{subsec:efficiency}

We evaluate the efficiency of \system{} using large-scale \textbf{Dataset III}. In particular, we measure the latency for the following three tasks: (1) the \sfg{} generation time for a single contract, (2) the \sfg{} embedding time from the \sfg{}, and (3) the vector searching time in the dataset. Our experiments are conducted on a server equipped with Intel(R) Core(TM) i7-9700 CPU running at 3.00GHz, 32 GB memory and 1TB SSD.

The latency associated with the first two tasks reflects the time required to analyze a new contract, whereas the latency of the third task reflects the time taken to search our database for similar contracts once the analysis of the new contract is complete. The evaluation results, illustrated in Figure~\ref{fig:dotvecdbtime}, provide insights into the performance of our system. Overall, the average time for constructing the \sfg{} is measured to be 26 seconds. In comparison, the average time for generating embeddings is 0.06 seconds, which can be considered negligible. This demonstrates the efficiency of our approach.
It is important to note that the EVM binaries have a size limit, typically no more than 24,576 bytes.
Consequently, the time taken for contract analysis does not increase indefinitely with the size of the bytecode. The query time increases linearly with the Data Volume. Therefore, even for a full-text search across all contracts currently deployed on six blockchains in \textbf{Dataset III}, the time can be limited to approximately 5 seconds.

\begin{tcolorbox}[size=title,colback=white]
\textbf{Answer to RQ3:} \system{} is efficient, the average time for constructing per contract's \sfg{}s is 26 seconds. The search time across all contracts on six blockchains can be limited to 5 seconds. 
\end{tcolorbox}

\section{Real-world Case Study}
\label{sec:case}

\begin{table}[!tb]
\centering
\caption{The Esim search results and comparison with Etherscan for similar vulnerable code search.}
\label{tab:vulall}
\resizebox{.45\textwidth}{!}{
\begin{threeparttable}
\begin{tabular}{lcccc}
\hline
\multirow{3}{*}{Project}  & \multirow{3}{*}{Loss} & \multirow{2}{*}{\textbf{Esim}} & \multicolumn{2}{c}{Etherscan}  \\ \cline{4-5} 
  &   &  & High  & Medium  \\ \cline{3-5} 
    &   & TP/FP & TP/FP & TP/FP   \\ \hline
Hedgey Finance~\cite{Hedgey}   & 1.8M   & 2/0   & 2/0   & 0/0\\
Seneca~\cite{Seneca}  & 6M& 15/0 & 1/0   & 2/84    \\
Abracadabra~\cite{MIMSpell}   & 6.5M   & 15/0 & 2/0   & 4/99    \\
Socket~\cite{SocketAttack}     & 3.2M   & 6/0   & 3/0   & 3/0\\
Gamma Strategies~\cite{Gamma}& 6.5M   & 66/0 & 65/1  & 12/22   \\
Radiant Capital~\cite{Radiant} & 4.5M   & 135/0    & 36/0  & 13/338  \\
Floor Protocol~\cite{Floor}   & 1.6M   & 16/0 & 1/0   & 3/2\\
NFT Trader~\cite{NFTTrader}    & 3M& 3/0   & 2/0   & 0/0\\
KyberSwap~\cite{KyberSwapAttack}     & 46M    & 6/0   & 6/0~\tnote{1} & -\tnote{2}\\
Raft~\cite{Raft}      & 3.2M   & 2/0   & 2/0   & 0/0\\
OnyxProtocol~\cite{OnyxProtocol} & 2M& 34/0 & 10/0  & 14/337  \\
Platypus~\cite{Platypus}   & 2M& 18/0 & 6/0   & 8/26    \\
BH Token~\cite{BHToken}    & 1.3M   & 8/0   & 3/0   & 0/0\\
Stars Arena~\cite{StarsArena} & 3M& 6/0   & 1/0   & 4/0\\
Balancer~\cite{BalancerAttack} & 2M& 48/0 & 19/0  & 29/298  \\
Exactly Protocol~\cite{ExactlyAttack}    & 7M& 2/0   & 2/0   & 2/194 \\ \hline
\textbf{Total}   & - & \textbf{382/0}  & \textbf{161/1}  &  \textbf{94/1400}  \\ \hline
\end{tabular}%

\begin{tablenotes}
\item[1] Contract addresses are consolidated based on factory contracts.
\item[2] Etherscan's results exceed the display limit and contain excessive false positives.
% \item[3] It's a fork of Compound V2~\cite{Compound} project.
% \item[4] The two positive samples are the results after the bug is fixed.
\end{tablenotes}

\end{threeparttable}
}
\vspace{-1em}
\end{table}

% Floor 4
An important application of similarity detection is to enable a quick response to zero-day attacks by identifying smart contracts with similar vulnerabilities, thereby mitigating the risk of subsequent one-day attacks. 
Unlike DeFi developers, third-party security researchers or companies, which may not have access to source codes, must rely on bytecode-level tools to perform such detection.

To evaluate the effectiveness of our system, we perform \system{} on DeFi attack incidents~\footnote{These incidents were sourced from a Web3 security company, BlockSec~\cite{BlockSec}.} that resulted in losses exceeding \$1 million between May 2023 and May 2024. We used the vulnerable codes in these DeFi attacks as benchmarks.
% and applied \system{} to \textbf{Dataset III} to search for contracts with similar vulnerabilities.
% To broaden the range of vulnerabilities identified and ensure search accuracy, we implemented a dynamic similarity threshold. 
% Specifically, the threshold was set at the point where the similarity distance interval exhibited a significant increase.

After searching, three engineers conducted a manual review to assess the accuracy. 
The review focused on three key aspects: (i) input parameters, including the number and type of parameters at the entry point; (ii) the main control flow, including \code{call}, \code{staticcall}, and \code{delegatecall}; and (iii) output logs, \ie, \code{log} events. 
After each engineer cast their votes, a decision was ultimately made based on the principle of the minority obeying the majority.
We compared \system{} with Etherscan~\cite{Etherscan}, the most widely used industrial similarity detection tool. Etherscan offers three levels of similarity search: \textit{high}, \textit{medium}, and \textit{low}. In our experiments, low-similarity search produced a false positive (FP) rate exceeding 95\%, making it ineffective for locating actual vulnerabilities. Therefore, we focused on high- and medium-similarity search results for comparison. 

After manual review, we found all high-similarity search results to be true positives. As shown in Table~\ref{tab:vulall}, \system{} identified 382 contracts with similar vulnerabilities, achieving zero false positives and significantly outperforming Etherscan. Although Etherscan's high-level results were accurate, the number of correct matches was lower, with only 40\% of the valid results found by \system{} (161 vs. 382). 
For medium-similarity results, Etherscan's correct matches increased, but the FP rate rose to 93.3\%. Even with this high FP rate, Etherscan’s total correct results (255) remained lower than \system{}'s (382).
The superior accuracy and higher number of results demonstrate the effectiveness of \system{} in identifying similar vulnerable code, achieving an optimal balance between search quantity and precision.

\section{Discussion}
\label{sec:discussion}

% \sfg{} is a tailored representation for EVM bytecode designed to capture the DeFi semantics of the code while remaining stable against \solc{} the basic block reuse strategy and version updates. We explore its limitations and potential improvements in detail below.

\noindent\textbf{DeFi attack response.}
In Section~\ref{sec:case}, we applied \system{} to locate vulnerability. 
After manually reviewing the evaluation results, we found that most of the funds in the vulnerable contracts had already been drained by attackers due to the time lag between our experiments and the attack events. 
% Our findings suggest that once a vulnerability is exploited, similar vulnerable contracts are likely to be targeted soon. 
To mitigate this, \system{} should be integrated with a real-time attack detection system. Upon detecting an attack, \system{} can swiftly identify potential victims and aid in preventive measures (the average search time of \system{} is less than 2s).

% \noindent\textbf{Static analysis limitations.} 
% Most EVM contracts are compiled using the widely adopted Solidity compiler, \solc{}, and current static analysis techniques rely on \solc{}'s characteristics. However, contracts can also be written directly in EVM bytecode, bypassing \solc{}'s features, which poses a challenge for static analysis. 
% In addition, bytecode obfuscation techniques~\cite{bytecodeobfuscation} further complicate the analysis, potentially leading to failures in our representation.

% \noindent\textbf{Adaptability to EVM development.} 
% Our representations are based on the current EVM design, emphasizing stable opcode selection and execution flow routes. However, as the EVM is rapidly evolving, new mechanisms such as jump methods, opcodes, and storage features may emerge. Therefore, we cannot guarantee the validity of our representations in future EVM versions. However, we believe that by adapting to new mechanisms, our approach can maintain its relevance and guiding significance.

\noindent\textbf{Extensibility to other languages.}
While this study focuses exclusively on EVM bytecode representation and evaluation, our approach is not limited to the EVM. 
The core concept is to recognize the differing significance of various instructions (\ie, stack-based and on-chain storage instructions in EVM) to simplify the complex control flow graph (CFG) using instruction sets of varying importance. 
We believe that this methodology can be extended to other programming languages and scenarios, enabling a tailored simplification of CFG complexity for specific contexts.
\section{Related Work}
\label{sec:relatedwork}

% \noindent 
% \textbf{Static Analysis of Binary.} Many researchers have already worked on the problem of bug search in binaries and made a great contribution to this direction. Vandal~\cite{brent2018vandal} and Gigahorse \cite{grechGigahorseThoroughDeclarative2019}\cite{grechElipmocAdvancedDecompilation2022} provide an efficient and scalable analysis framework, allowing researchers to write detectors in datalog language based on them. It is adapted by many security tools, such as Madmax~\cite{grech_madmax_2020}. Ethersolve~\cite{contro2021ethersolve} is a tool generating CFG for EVM bytecode by using symbols to emulate stack. Mythril is a security analysis tool for EVM bytecode. It uses symbolic execution, SMT solving, and taint analysis to detect a variety of security vulnerabilities. Panoramix~\cite{panormix} is used by Etherscan, using symbolic execution to explore paths and reconstruct a high-level representation.

\noindent\textbf{Binary Function Representation Problem}
The issue of binary function similarity is a widespread and extensively researched problem. There are various methods available, ranging from straightforward to intricate. Binary function representation can be categorized into two main directions: instruction stream based studies~\cite{zuo2018neural,pei2022learning,wang2022jtrans}, and control flow graph-based studies (\eg, CFG~\cite{huang2017binsequence,nouh2017binsign}, ACFG~\cite{feng2016scalable,xu2017neural,marcelli2022machine}).
Prior research has primarily concentrated on the analysis of C/C++ and JVM binaries. However, these kinds of binary representations face new challenges in EVM bytecode. The EVM bytecode has distinct features, \ie, low level, piecemeal \& heavily reused blocks, highly-diverse \solc{} update \& options. 
In this study, we propose a novel graph-format feature, \sfg{}, that can effectively present the EVM bytecode DeFi semantics. 

\noindent \textbf{Machine Learning in Binary Code Similarity.}
With the development of artificial intelligence techniques, machine learning with more powerful feature extraction capabilities is applied to binary code similarity. For example, GNN is widely used due to its ability to effectively capture structural information~\cite{gao2018vulseeker,li2019graph,xu2017neural}. Some studies also leverage advanced models to learn the semantics of representation~\cite{he2023finer}, such as GGNN and GIN~\cite{li2019graph,he2024code,he2023msdroid}. Marcelli et al.~\cite{marcelli2022machine} analyze the machine learning behavior in binary code similarity and prove the GNN effect. 
In our study, our representation is mixed with control flow and data flow. Inspired by that, we leverage the heterogeneous graph embedding model to combine the control flow information and data flow information and perform similarity detection.
\section{Conclusion}
\label{sec:conclusion}

In recent years, DeFi is undergoing substantial growth. However, the blockchain ecosystem faces significant security challenges due to extensive code reuse and the scarcity of open-source code, leading to issues such as the proliferation of vulnerable code.
% This highlights the urgent need for effective similarity detection methods for EVM bytecode.
In this study, we propose a novel EVM bytecode representation called the Stable-Semantic Graph (SSG). We developed a prototype, \system{}, which demonstrates high accuracy in stable semantic construction, achieving a 100\% F1-score for control flow and 95.16\% for data flow. Its similarity performance reached 96.3\% AUC, surpassing traditional studies and achieving SOTA performance.
Subsequently, we conducted large-scale case studies analyzing 2,675,573 contracts across 6 EVM-compatible blockchains. Our evaluation shows that \system{} outperforms the current most popular industry tool, Etherscan, in vulnerability code search.

% \section*{Data Availability Statement}
% The prototype of the proposed \system{} system is available in an anonymous repository at \url{https://anonymous.4open.science/r/esim-CF8D} and will be made publicly available after acceptance. The datasets for \system{} can be obtained from the corresponding author upon reasonable request.

\bibliographystyle{IEEEtran}
\bibliography{sample-base}

% \newpage
% \appendix
% \section{Appendix}
% The appendix includes one figure and two tables of information. 
% Figure~\ref{fig:tsne} is the 2D embeddings of different source functions in Section~\ref{subsec:modelevaluate}.
% Table~\ref{tab:uniswapv3all} provides detailed information about the contract addresses that infringe on the Uniswap v3 protocol in Section~\ref{subsec:copyright}. Table~\ref{tab:rugall} contains specific details about the malicious family clustering in Section~\ref{subsec:rugpull}.

% \begin{figure}[!h]
% \centering
% 	\includegraphics[width=.4\textwidth]{Fig/ESIM-TSNE.drawio.pdf}
% 	\caption{Visualizing the embeddings of the different functions using t-SNE. Each color indicates one function.}
% 	\label{fig:tsne}
%         \vspace{-1em}
% \end{figure} 

% \input{Tab/uniswapv3all}

% \input{Tab/RUGall}

\end{document}